\documentclass[11pt]{article}
\usepackage{amstext, amsmath,latexsym,amsbsy,amssymb,amsmath}
\newenvironment{dedication}
  {
   \thispagestyle{empty}
   \vspace*{\stretch{1}}
   \itshape             
   \raggedleft          
  }
  {\par 
   \vspace{\stretch{3}} 
   \clearpage           
  }

\usepackage[text={5.3in,7.2in},centering,letterpaper]{geometry}

\usepackage{enumitem}

\numberwithin{equation}{section}

\newcommand{\QED}{\hspace*{\fill}\rule{2.5mm}{2.5mm}}

\usepackage{amssymb}\usepackage{graphicx}
\usepackage{tikz}

\usepackage{color}
\newcommand\qed{\hfill$\sqcap\kern-7.5pt\hbox{$\sqcup$}$}

\newcommand{\beqn}{\begin{equation}}
\newcommand{\eeqn}{\end{equation}}
\newcommand{\bear}{\begin{eqnarray}}
\newcommand{\eear}{\end{eqnarray}}
\newcommand{\bean}{\begin{eqnarray*}}
\newcommand{\eean}{\end{eqnarray*}}

\allowdisplaybreaks

\begin{document}
\title{On the Wave Turbulence Theory for the Nonlinear Schr\"odinger Equation with Random Potentials}


\author{Sergey Nazarenko\footnotemark[1] \and Avy Soffer\footnotemark[2] \and Minh-Binh Tran\footnotemark[3]
}

\renewcommand{\thefootnote}{\fnsymbol{footnote}}

\footnotetext[1]{
Universit\'e C\^ote d’Azur,  CNRS, Institut de Physique de Nice,
 Parc Valrose, 06108 Nice, Franc.\\Email: sergey.nazarenko@unice.fr
}

\footnotetext[2]{Mathematics Department, Rutgers University, New Brunswick, NJ 08903 USA.\\Email: soffer@math.rutgers.edu
}
\footnotetext[3]{Department of Mathematics, Southern Methodist University, Dallas, Texas, TX 75275, USA. \\Email: minhbinht@mail.smu.edu
}
\maketitle

{\dedication{~~~~~~~~~~~~~~~~~~~~~Dedicated to the Memory of Shmuel Fishman}}

\begin{abstract}
We derive a new kinetic and a porous medium equations from the nonlinear Schr\"odinger equation with random potentials.  The kinetic equation has a very similar form with the 4-wave turbulence kinetic equation in the wave turbulence theory. Moreover, we construct a class of self-similar solutions for the porous medium equation. These solutions spread infinitely as time goes to infinity  and this fact  answers the ``weak turbulence'' question for the  nonlinear Schr\"odinger equation with random potentials positively. We also derive  Ohm's law for the porous medium equation.
 \end{abstract}

{\bf Keyword:}   Wave Turbulence Theory, Nonlinear Schr\"odinger Equation with Random Potentials, 4-Wave Kinetic Turbulence Equation, Ohm's Law, Porous Medium Equation.

\tableofcontents
\section{Introduction}

The nonlinear Schr\"odinger equation (NLSE) with random potentials is a fundamental problem in both mathematical and physical research. Although there have been extensive mathematically rigorous, analytical and numerical results, several elementary properties of the dynamics of the solutions are still not known. The resolution of the problem plays a central role in understanding several physical phenomena in chaos and nonlinear physics. The NLSE  with random potentials is written as follows,
\begin{equation}
\label{NLSRandom}
\begin{aligned}
i\partial_t \Psi(x,t) \  = & \ H_0\Psi(x,t) \ +  \  \epsilon |\Psi(x,t)|^2\Psi(x,t) \\
= & \ - \Delta \Psi(x,t)\ + \ V_x \Psi(x,t) \ + \ \epsilon |\Psi(x,t)|^2\Psi(x,t) ,
\end{aligned}
\end{equation}
where  $V_x$ is a random function.

On a one-dimensional lattice,
$$x\in \xi\mathbb{Z}:= \{ \xi n, \ \ n\in \mathbb{Z}\},$$  the lattice version of the above equation can be written as
\begin{equation}
\label{NLSRandom1}
\begin{aligned}
i\partial_t \Psi_x \  = & \ \frac{1}{\xi^2}[\Psi_{x+\xi}+\Psi_{x-\xi}-2\Psi_{x}] \ +  \  \epsilon |\Psi_x|^2\Psi_x + \ V_x \Psi_x,
\end{aligned}
\end{equation}
where $V_x$ is  a collection of i.d.d. random variables uniformly distributed in the
interval $[-\omega/2,\omega/2]$.
In this paper,  we will focus on the so-called ``weak turbulence'' question about the dynamics of the solution in large time:
\begin{equation}
\label{Question}
\begin{aligned}
& \mbox{\it Will a small nonlinearity spread the  solution  over distances much greater than }\\
& \mbox{\it the linear  system  does for large times for an  initial condition localized}\\
& \mbox{\it in space and frequency?}
\end{aligned}
\end{equation}
 This question is still open despite several efforts \cite{bourgain2009mathematical,FaouGermainHani:TWN:2016,hani2014long,pikovsky2008destruction,wang2009long,flach2009universal,skokos2009delocalization,fishman2009perturbation,krivolapov2010numerical,fishman2012nonlinear,Flach2015}.  The resolution of this question may shed lights on many nonlinear problems, such as the famous Fermi-Pasta-Ulam (FPU) problem \cite{berman2005fermi,campbell2005introduction}. We also refer to \cite{Flach:SNS:2016,mithun2018weakly,vakulchyk2019wave} for recent numerical  works on the microcanonical Gross-Pitaevskii (also known as the semiclassical Bose-Hubbard) lattice model dynamics.

 A convincing evidence for delocalisation by nonlinearity has been gained by numerical experiment, and several intuitive arguments and phenomenological descriptions were suggested. However, there still remains a need of a more solid theory based on more rigorous and systematic derivations with clearly spelled out starting assumptions.
 Our approach is then to use the wave turbulence approach \cite{Nazarenko:2011:WT,zakharov2012kolmogorov,lvov2004noisy,balk1990physical,galtier2000weak} to derive a kinetic equation from the 1D latice NLSE with random potentials \eqref{NLSRandom}.
Superficially, the derived kinetic equation has a  similar form with the four-wave turbulence kinetic equation and the quantum Boltzmann equation \cite{Nazarenko:2011:WT,NewellNazarenko:2001:WTA,SofferBinh2,SofferBinh1,JinBinh,ReichlTran,PomeauBinh,CraciunBinh}. However, there is no conservation of momentum due to the localisation in space of the linear modes, and the modes are parametrised by their location on the lattice rather by their momentum-space location.
On the other hand, it is the localised nature of the linear modes and the fact that the interactions happen only locally in physical space that allow us, in the first order approximation, transform the kinetic equation into a nonlinear diffusion equation -- a.k.a. the porous medium equation.

For the porous medium equation, we construct a self-similar  solution that spreads infinitely for large times. This fact  answers positively the ``weak turbulence'' question \eqref{Question}. We also find a class of steady state solutions to the equation, that leads to a new nonlinear  Ohm's law. This result implies that small nonlinearity breaks down the insulator property of the latice and turns it into a nonlinear conductor.


\section{Derivation of the kinetic equation}

There have been several approaches to deriving wave turbulence in previous literature.
For our purposes, most suited is  the wave turbulence technique of book \cite{Nazarenko:2011:WT}
(see also the original papers \cite{CHOI2004,lvov2004noisy,CHOI2005,CHOI2005a}). This approach is based on an explicit formulation of statistical properties
of waves by introducing ''the random phase and amplitude"  fields.
However, due to the localisation of the linear modes in presence of random
potential, we will have to adopt an extra element previously developed in \cite{rivkind2011eigenvalue}, assuming Wick's type behaviour of Nth order correlations of the linear problem,
 the so-called quasi-free field assumption.

\subsection{Dynamical equations for the mode amplitudes}

Let us represent the
the wavefunction in terms of the  eigenvalues $\mathcal{E}_j$ and
eigenvectors $\psi_j(x)$  of the Hamiltonian $H_0$:
$$\Psi(x,t)  \ = \ \sum_{j\in\mathbb{Z}}c_j(t)e^{-i\mathcal{E}_j t}\psi_j(x).$$

Substituting the above expansion into \eqref{NLSRandom} and removing all the oscillating linear terms, we find
\begin{equation}
\label{ODEsystem1}
\partial_t c_j(t)  \ =  \ \epsilon \sum_{l,m,n\in\mathbb{Z}}V_{jl}^{mn}c_l^*c_mc_ne^{i(\mathcal{E}_n-\mathcal{E}_l+\mathcal{E}_m-\mathcal{E}_j)t}\ =:  \ \epsilon \sum_{l,m,n\in\mathbb{Z}}V_{jl}^{mn}c_l^*c_mc_ne^{i\mathcal{E}^{mn}_{lj}t},
\end{equation}
in which
\begin{equation}
\label{Vmnjl}
V^{mn}_{jl}  \ =  \  \sum_{x\in\xi\mathbb{Z}}\psi_l^*(x)\psi_m(x)\psi_n(x)\psi_j^*(x)
 = V^{jl*}_{mn}.
\end{equation}

For further derivation, we need to remove the diagonal terms, which include terms satisfying $(n,m)=(l,j)$ and $(n,m)=(j,l)$. These terms can be expressed as follows,
$$\sum_{(n,m)=(l,j), (n,m)=(j,l)}V_{jl}^{mn}c_l^*c_mc_ne^{i(\mathcal{E}_n-\mathcal{E}_l+\mathcal{E}_m-\mathcal{E}_j)t}\  =\ 2c_j\sum_{n\in\mathbb{Z}}V_{jn}^{jn}|c_n|^2 \ =: \ E_{NL}c_j.$$

We then absorb these terms by defining the energy normalization

\begin{equation}
E \ := \  \mathcal{E} \ +  E_{NL},
\end{equation}
that gives
\begin{equation}
\label{ODEsystem}
\partial_t c_j(t)  \ =  \ \epsilon \sum^\prime_{l,m,n\in\mathbb{Z}}V_{jl}^{mn}c_l^*c_mc_ne^{i({E}_n-{E}_l+{E}_m-{E}_j)t} =:
\ \epsilon \sum^\prime_{l,m,n\in\mathbb{Z}}V_{jl}^{mn}c_l^*c_mc_ne^{i{E}_{mn}^{jl}t},
\end{equation}
where $\sum^\prime_{l,m,n\in\mathbb{Z}}$ denotes the sum in which the diagonal terms $(n,m)=(l,j)$ and $(n,m)=(j,l)$ are excluded.

So far we have not made any approximations and our ODE system (\ref{ODEsystem})
is equivalent to the original equation (\ref{NLSRandom1}).

\subsection{Weak nonlinearity expansion}

Let us now introduce the intermediate time $T$
\begin{equation}
\label{IntermediateTime}
\frac{2\pi}{E_j}  \ll \ T \ll \ \frac{2\pi}{E_j\epsilon^2}.
\end{equation}
For T in this range our approximations will make sense for most potentials (probability close to $1$).
Under the assumption that $\epsilon$ is very small, we can expand the coefficient $c_j(T)$ as
\begin{equation}
\label{ExpansionCJ}
c_j(T)  \ = \ c_j^{(0)}(T) \ + \ \epsilon c_j^{(1)}(T) \ + \ \epsilon^2 c_j^{(2)} (T) \ +  \ \cdots
\end{equation}

Inserting the expansion \eqref{ExpansionCJ} into \eqref{ODEsystem}, yields a new system of equation for $c_j^{(r)}$ that we describe below.

For $r=0$, the problem is linear, and we have the following equation
\begin{equation}
\label{EqCJ0a}
\partial_t c_j^{(0)} \ = \ 0,
\end{equation}
that implies
\begin{equation}
\label{EqCJ0}
 c_j^{(0)}(T) \ = \ c_j^{(0)}(0).
\end{equation}

For $r=1$, we find
\begin{equation}
\label{EqCJ1a}
i\partial_t c_j^{(1)} \ = \ \sum^\prime_{m,n,l\in\mathbb{Z}}V^{mn}_{lj}e^{iE^{mn}_{lj}t}c^{(0)}_mc^{(0)}_nc^{(0)*}_l
\end{equation}
which yields
\begin{equation}
\label{EqCJ1}
c_j^{(1)} \ = \ -i\sum^\prime_{m,n,l\in\mathbb{Z}}V^{mn}_{lj}\Delta_T({E^{mn}_{lj}})c^{(0)}_mc^{(0)}_nc^{(0)*}_l,
\end{equation}
where
$$\Delta_T({E^{mn}_{lj}}) \ = \ \int_0^Te^{iE^{mn}_{lj}t}dt \ = \ \frac{e^{iE^{mn}_{lj}T}-1}{iE^{mn}_{lj}}.$$

%
For $r=2$, the following equation can be obtained,
\begin{equation}
\label{EqCJ2a}
\begin{aligned}
i\partial_t c_j^{(2)} \ = & \ \sum^\prime_{m,n,l\in\mathbb{Z}}V^{mn}_{lj}e^{iE^{mn}_{lj}t}(2c^{(1)}_mc^{(0)}_nc^{(0)*}_l
+c^{(0)}_mc^{(0)}_nc^{(1)*}_l)\\
\ = & \ \ -i\sum^\prime_{m,n,l\in\mathbb{Z}}V^{mn}_{lj}e^{iE^{mn}_{lj}t}\Big(2c^{(0)}_nc^{(0)*}_l\sum_{\lambda,\mu,\nu\in\mathbb{Z}}V_{\lambda m}^{\mu\nu}\Delta_T(E_{\lambda m}^{\mu\nu})
c^{(0)}_\mu c^{(0)}_\nu c^{(0)*}_\lambda\\
&\ - c^{(0)}_mc^{(0)}_n\sum^\prime_{\lambda,\mu,\nu\in\mathbb{Z}}V_{\lambda l}^{\mu\nu *}\Delta^*_T(E_{\lambda l}^{\mu\nu})c^{(0)*}_\mu c^{(0)*}_\nu c^{(0)}_\lambda\Big),
\end{aligned}
\end{equation}
which yields
\begin{equation}
\label{EqCJ2}
\begin{aligned}
c_j^{(2)} \ = & \ \ -\sum^\prime_{m,n,l,\lambda,\mu,\nu\in\mathbb{Z}}V^{mn}_{lj}\Big[2c^{(0)}_nc^{(0)*}_lc^{(0)}_\mu c^{(0)}_\nu c^{(0)*}_\lambda V_{\lambda m}^{\mu\nu}\Gamma(E_{lj}^{mn},E_{\lambda m}^{\mu\nu})
\\
&\ - c^{(0)}_mc^{(0)}_nc^{(0)*}_\mu c^{(0)*}_\nu c^{(0)}_\lambda V_{\lambda l}^{\mu\nu *}\Gamma(E_{lj}^{mn},E^{\lambda l}_{\mu\nu})\Big],
\end{aligned}
\end{equation}
where
$$\Gamma(x,y) \ = \ \int_0^Te^{ixt}\Delta_t(y)dt.$$

Now, let us try to understand the spectrum by developing
\begin{equation}\label{Density}
\begin{aligned}
N_j(T) \ = & \ \langle |c_j(T) |\rangle ^2 \\
\ = & \ \langle |c_j^{(0)}(T) \ + \ \epsilon c_j^{(1)}(T) \ + \ \epsilon^2 c_j^{(2)}(T) \ + \cdots |^2\rangle\\
\ = & \ \langle |c_j^{(0)}(T)| ^2 \rangle \ + \ \epsilon\langle c_j^{(1)}(T)c_j^{(0)*}(T)\ +\  c.c. \rangle \ + \ \epsilon^2\langle |c_j^{(1)}(T)|^2\rangle \\
\ & \ +\ \epsilon^2\langle c_j^{(2)}(T)c_j^{(0)*}(T) \ + \ c.c.\rangle.
\end{aligned}
\end{equation}




\subsection{Statistical averaging}

Fundamentally, stochasticity of our system arises from randomness of the potentials at each lattice site.
Even if the initial mode amplitudes $c_j$ are deterministic, they will become random at a later time.
Moreover, it is natural to assume that $c_j$'s will become statistically independent at each site $j$ and that
their phases will become random.

Thus, let us assume that at the time $t=0$ such randomness is already ensured by the preceding evolution at $t<0$.
Specifically,  let us make the following assumptions:
\begin{itemize}
\item Assumption 1: Phase randomness. This is a standard wave turbulence assumption.  We assume that the phases of $c^{(0)}_j$ are random and, therefore, we can use Wick's pairing, which says that non-zero contributions only arise in paring $c_j^{(0)}$ and $c_j^{(0)*}$. That means the first order term in $\epsilon$ is $0$
$$\langle c_j^{(0)*}c_j^{(1)}\rangle \  = \ -i\sum^\prime_{m,n,l\in\mathbb{Z}}\langle V_{lj}^{mn}c^{(0)}_mc^{(0)}_nc^{(0)*}_lc_j^{(0)*}\Delta (E_{ij}^{mn})\rangle=0.$$

This result holds
because the diagonal terms with $(m,l)=(n,j)$ and $(m,n)=(l,j)$ are excluded from the sum.

The second order terms in  $\epsilon$ can be written as
\begin{equation}\label{QuasiFreeo}
\begin{aligned}
\langle
|c_j^{(1)}|^2
\rangle = &\sum^\prime_{l,m,n,\lambda,\mu,\nu\in\mathbb{Z}} \langle V^{mn}_{lj}\Delta_T({E^{mn}_{lj}})c^{(0)}_mc^{(0)}_nc^{(0)*}_l
V^{\mu\nu *}_{\lambda j}\Delta^*_T({E^{\mu\nu}_{\lambda j}})c^{(0)*}_\mu c^{(0)*}_ \nu c^{(0)}_\lambda \rangle
\
\\
=& 2 \ \sum^\prime_{m,n,l\in\mathbb{Z}}\langle |V_{lj}^{mn}|^2 |c_m^{(0)}|^2 |c_n^{(0)}|^2|c_l^{(0)}|^2 |\Delta (E_{lj}^{mn})|^2 \rangle
\end{aligned}
\end{equation}
and
\begin{equation}\label{QuasiFree1}
\begin{aligned}
\langle c_j^{(2)}c_j^{(0)*}\rangle \ = &\  -\sum^\prime_{m,n,l,\lambda,\mu,\nu\in\mathbb{Z}}\Big\langle V^{mn}_{lj}\Big[2c^{(0)}_n c^{(0)}_\mu c^{(0)}_\nu c^{(0)*}_l  c^{(0)*}_\lambda c_j^{(0)*} V_{\lambda m}^{\mu\nu}\Gamma(E_{lj}^{mn},E_{\lambda m}^{\mu\nu})
\\
&\ - c^{(0)}_mc^{(0)}_n c^{(0)}_\lambda c^{(0)*}_\mu c^{(0)*}_\nu  c_j^{(0)*} V_{\lambda l}^{\mu\nu *}\Gamma(E_{lj}^{mn},E^{\lambda l}_{\mu\nu})\Big]\Big\rangle\\
\ = &\  -2\sum^\prime_{m,n,l\in\mathbb{Z}}\Big\langle |V^{mn}_{lj}|^2 \Gamma(E_{lj}^{mn},E_{ mn}^{lj}) \left[2|c^{(0)}_l|^2|c^{(0)}_n|^2 |c^{(0)}_j|^2 -
|c^{(0)}_m|^2|c^{(0)}_n|^2 |c^{(0)}_j|^2 \right]\Big\rangle
.
\end{aligned}
\end{equation}
\item Assumption 2: Amplitude averaging -- ``quasi-free field'' assumption. Here we will assume that
the mode amplitudes at each site are statistically independent. Moreover, we will assume that
these amplitudes are independent from $|V_{lj}^{mn}|^2  |\Delta (E_{lj}^{mn})|^2$ and $|V^{mn}_{lj}|^2 \Gamma(E_{lj}^{mn},E_{ mn}^{lj})$ (complicated functions of the random potentials which are fixed for each realisation).  This assumption leads to
\begin{equation}\label{QuasiFree1}
\begin{aligned}
 &\langle |V_{lj}^{mn}|^2 |c_m^{(0)}|^2 |c_n^{(0)}|^2|c_l^{(0)}|^2 |\Delta (E_{lj}^{mn})|^2 \rangle
\\
= \ &\ \langle|V_{lj}^{mn}|^2  |\Delta (E_{lj}^{mn})|^2 \rangle \langle  c_m^{(0)}|^2\rangle \langle |c_n^{(0)}|^2\rangle\langle |c_l^{(0)}|^2 \rangle
\end{aligned}
\end{equation}
and
\begin{equation}\label{QuasiFree2}
\begin{aligned}
  \left\langle |V^{mn}_{lj}|^2 \Gamma(E_{lj}^{mn},E_{ mn}^{lj}) \left[2|c^{(0)}_l|^2|c^{(0)}_n|^2 |c^{(0)}_j|^2 -
|c^{(0)}_m|^2|c^{(0)}_n|^2 |c^{(0)}_j|^2 \right]\right \rangle = \\
\left\langle |V^{mn}_{lj}|^2 \Gamma(E_{lj}^{mn},E_{ mn}^{lj}) \right\rangle \left[2\langle|c^{(0)}_l|^2\rangle\langle|c^{(0)}_n|^2 \rangle\langle|c^{(0)}_j|^2 \rangle-
\langle|c^{(0)}_m|^2\rangle\langle|c^{(0)}_n|^2 \rangle\langle|c^{(0)}_j|^2 \rangle\right]
\end{aligned}
\end{equation}
\end{itemize}
Note that assumption about random statistically independent wave amplitudes is one of the key
elements of the wave turbulence technique of book \cite{Nazarenko:2011:WT}.
However, the quasi-free field assumption goes beyond this assumption by additionally
assuming that the wave amplitudes get decorrelated from  $V_{lj}^{mn}$ and $E_{lj}^{mn}$
 which
contains the original randomness source via the random nature of the linear eigenmodes
and eigenvalues.

\subsection{Four-wave kinetic equation}

According to the inequality \eqref{IntermediateTime}, the small nonlinearity limit
corresponds to  $T\to \infty$, and in this case we know that
\begin{equation}
\label{d1}
|\Delta (E_{lj}^{mn})|^2 \to 2 \pi T  \delta (E_{lj}^{mn}) , \quad \hbox{and}
\quad \Gamma(E_{lj}^{mn},E_{ mn}^{lj}) + \Gamma^*(E_{lj}^{mn},E_{ mn}^{lj})
\to 2 \pi T  \delta (E_{lj}^{mn}).
\end{equation}

{\bf Remark 1.}
In continuous models without potential, the large-box limit is taken before the weak
nonlinearity limit. This makes the momentum space continuous, and it is only for continuous
space that Dirac delta function can be introduced. In our discrete NLS, the "box" is
infinite from start (infinite lattice), but the linear modes are still discrete.
However, the Dirac delta function in the above expressions is well defined because
of the averaging operator, which can be viewed as an integral over the possible
realisations of $E_j$ which form a continuous set (since the set of possible values of potentials
is continuous).

{\bf Remark 2.}
In continuous NLS without potential, the dispersion relation does not allow four-wave resonances
in 1D. Thus we can expect that the above expressions will become null if
we make the lattice spacing in our discrete  system or the maximum potential $\omega$ too small.
In this case we expect the six-wave process to be more effective than the four-wave process (see more
about this later in this paper).

  Let us denote
\begin{equation}
\label{Kernel}
K(l-j,m-j,n-j) \ := \  4 \pi  \epsilon^2  \langle |V_{lj}^{mn}|^2  \delta (E_{lj}^{mn}) \rangle,
\end{equation}
where we took into account the fact that all the lattice sites are equivalent and, therefore, $K$ may
depend of the difference of the site indices only.

Substituting \eqref{QuasiFree1} and \eqref{QuasiFree2} into \eqref{Density}
and taking into account \eqref{d1} and that $(N_j(T) - N_j(0))/T \approx \dot N_j(T)$, we have
the following kinetic equation,
  \begin{equation}
\label{Kinetic0}\begin{aligned}
\dot{N}_j \ = & \  \sum^\prime_{m,n,l\in\mathbb{Z}} K(l-j,m-j,n-j)(N_lN_mN_n+N_nN_mN_j-N_jN_nN_l-N_lN_jN_m),
 \end{aligned}
\end{equation}
This kinetic equation has a   form similar to the 4-wave turbulence kinetic equation \cite{Nazarenko:2011:WT,zakharov2012kolmogorov}
 and the quantum Boltzmann equation (cf. \cite{Nazarenko:2011:WT,germain2017optimal,NewellNazarenko:2001:WTA,SofferBinh2,SofferBinh1,zakharov2012kolmogorov}).
 In particular, all these systems conserve the total mass, $\sum_j N_j$ and the characteristic evolution time scales as $1/\epsilon^2$.
 However,  there are important differences.
 Firstly, because the modes are localised rather than being monochromatic waves as in usual wave turbulence, there is no conservation of momentum.
 Secondly, also due to the localisation, the modes are parametrised by their sites at the lattice rather than by their momenta, as in usual wave turbulence. In fact, the linear mode in our case are equivalent to each other in terms of their momentum content.
 Thirdly, for strong fluctuating potentials with amplitudes $\omega \sim 1$, the eigenfunctions
$\psi_j$ are strongly localised around respective $j$. This means that in this case $K(l-j,m-j,n-j)$
 is strongly peak near $l=m=m=j$ and it is not possible to take a continuous limit in the kinetic equation.

On the other hand, the localisation is less strong for weaker $\omega $, and the width of $K(l-j,m-j,n-j)$ is greater.
In the case when such a width is significantly greater than the lattice spacing, one can pass to the continuous limit and write
\begin{equation}
\label{Kinetic1}\begin{aligned}
\dot{N}_j \ = \iiint_{\mathbb{R}^3}K(l-j,m-j,n-j)N_jN_lN_mN_n(N_j^{-1}+N_l^{-1} -N_m^{-1} -N_n^{-1})dmdndl,\end{aligned}
\end{equation}

\section{The porous medium equation}

Let us now exploit the property of locality of the kernel in the kinetic equation arising
from the localisation of the linear eigenmodes. This will allow us to reduce the integro-differential
kinetic equation to a simpler nonlinear diffusion equation. This will also allow us to obtain solutions
corresponding to delocalisation.

\subsection{Derivation of the porous medium equation}

Let us now consider the case where the modes are not too localised so that the continuous kinetic equation works, but, at the same time, the width of $K$ is much less than the characteristic length of dependence of $N_j$ on $j$.
Let us multiply equation \eqref{Kinetic1} by an arbitrary function $f(j)$ and integrate over $j$.
Split the result in 4 equal parts and the last three parts change variables as, respectively:
$ j \leftrightarrow l, m \leftrightarrow n$; $ j \leftrightarrow m, l \leftrightarrow n$; $ j \leftrightarrow n, l \leftrightarrow m$. These transformations leave function $K$ unchanged, so we have:
\begin{equation}
\label{Kinetic2}\begin{aligned}
\int \dot{N}_j f_j dj \ = \frac 1 4 \iiiint K(l-j,m-j,n-j)N_lN_mN_n N_j \times \\
( N_j^{-1} +  N_l^{-1} - N_m^{-1} -N_n^{-1} )
( f_j +  f_l - f_m -f_n ) dmdndldj.  \end{aligned}
\end{equation}
Taylor expanding the expressions in both brackets to the leading order in small
$\tilde l = l-j$, $\tilde m = m-j$, $\tilde n = n-j$, and writing $N_lN_mN_n N_j \approx N_j^4$ we
have:
\begin{equation}
\label{Kinetic3}\begin{aligned}
\int \dot{N}_j f_j dj \ = \frac 1 4 \iiiint K(\tilde l, \tilde m, \tilde n)N_j^4
(\tilde l - \tilde m - \tilde n)^2
( N_j^{-1})'
f_j'  d\tilde md\tilde nd \tilde ldj,  \end{aligned}
\end{equation}
where prime denotes differentiation with respect to $j$.
Integrating by parts with respect to $j$ and rearranging, we get:
\begin{equation}
\label{Kinetic4}\begin{aligned}
\int \dot{N}_j f_j dj \ = - \frac 1 4 \iiiint K(\tilde l, \tilde m, \tilde n) (\tilde l - \tilde m - \tilde n)^2
N_j^4 ( N_j^{-1})'
f_j d\tilde md\tilde nd \tilde ldj. \end{aligned}
\end{equation}
Since $f_j$ is arbitrary, we can drop $j$-integration on both sides.
Rearranging, we finally obtain:
\begin{equation}
\label{PorousMedium}
\dot N_j = {\cal D} \partial_{jj}N_j^3,
\end{equation}
where
\begin{equation}
\label{PorousMediumD}
{\cal D}  = \frac 1 {12} \iiint K(\tilde l, \tilde m, \tilde n) (\tilde l - \tilde m - \tilde n)^2
d\tilde md\tilde nd \tilde l.
\end{equation}

Equation \eqref{PorousMedium} is a nonlinear diffusion equation
(with the diffusion coefficient $3 {\cal D}  N_j^2$) which belongs to the class of
porous medium equations \cite{Vazquez:TPM:2007}.
Let us consider now  an extension of \eqref{PorousMedium}:
\begin{equation}
\label{PorousMedium2}
\partial_t N(t,k) =\partial_{kk}N^m(t,k), \ \ \  m>1,
\end{equation}
where we keep in mind that $m=3$ and $m=5$ correspond to the four-wave
and the six-wave systems respectively. (Later we will discuss the conditions under which the six-wave
dynamics occurs.)
Let us now consider solutions of this equation porous medium equation.

\subsection{Steady state solutions -  Ohm's law}
The porous medium equation has a conservation law form,
so we will use here an electricity terminology, i.e. we will call $N$ a charge density.
The steady state solution of the porous medium equation takes the form
\begin{equation}
\label{PorousMediumSteadyState}
\partial_{kk}N^m(k)  \ = \ 0.
\end{equation}

As a consequence
\begin{equation}
\label{PorousMediumSteadyState1}
\partial_{k}N^m(k)  \ = \ -J,
\end{equation}
where $J$ is a current constant.
Integrating once more, we get
$
N^m(k)  \ = \ A-Jk,
$
where $A$ is a constant, and hence
\begin{equation}
\label{PorousMediumSteadyState3}
N(k)  \ = \ [A-Jk]^\frac1m.
\end{equation}
As we see, for $A,J>0$ the charge density drops to zero at
a location $k=a = A/J$. Since $J$ is a $k$-independent constant,
one must put an "electrode" at $k=a$ that absorbs current $J$.

Let us now introduce a potential $\phi$ via
\begin{equation}
\label{PotentialEq}
\phi'' \ = \ N,
\end{equation}
which can be solved directly,
\begin{equation}
\label{Potential}
\phi \ = \ \frac{m^2}{(2m+1)(m+1)J^2}[A-Jk]^{\frac1m +2}  \ + \ Dk \ +  \ P.
\end{equation}
where $D$ and $P$ are some constants.
Since the potential is defined up to a constant only, we can
fix it by condition $\phi(a) = 0$, which gives
$
Da +P =0. $
In the other words, the electrode at $a$ is a "ground".
Also, $D$ has a meaning of a constant part of the electric
field. It is actually not observable in our problem, so we set
$D=P=0$ leading to $A=Ja$.

For "voltage", we find
\begin{equation}
\label{Phi0}\begin{aligned}
V = \phi(0) - \phi(a) \ =
\frac{m^2}{(2m+1)(m+1)J^2}A^{\frac1m +2} =
\frac{m^2 a^{\frac1m +2}}{(2m+1)(m+1)}J^{\frac1m} .
\end{aligned}
\end{equation}
This is an analog of Ohm's law describing the relation between the voltage and the
current. Note that in our case, the remaining traces of the localisation effect lead to nonlinearity of  Ohm's law.
However, since the current $J$ is finite for any $V>0$, the medium is conducting, i.e. the localisation (insulator) property is broken by the nonlinearity.

\subsection{Self-similar solutions}
Let us now consider another, more traditional approach to examining delocalisation -- time dependent evolution (spreading) of initially localised distributions. Essentially, most of the results of this sections were
previously obtained in \cite{tuck76,Kolovsky10,Flach2015}, and here we reproduce and summarise them for completeness of discussion.

Let us look for self-similar solutions of the first kind of the equation of \eqref{PorousMedium},
\begin{equation}
\label{SelfSimilar}
N(t,k) \ =  \ t^b f(\xi),
\end{equation}
where $\xi = k t^a, \; \; k,t > 0$, and $a$ and $b$ are some constants. For consistency of the formulation, we must satisfy
$2a = -1 +b(1-m)$.
The total mass conservation,
$\int N(t,k) dk =$const, gives the second condition: $a=b$. Thus,
\begin{equation}
\label{Xi}
\xi \  = \ k t^{-\frac1{(m+1)}}.
\end{equation}
The rate of spreading of initially localised distributions is usually measured by the evolution
of the standard deviation defined as $\sigma = \int k^2 N(k,t) dk.$ In our case we have
\begin{equation}
\label{sigma}
\sigma(t) = \int k^2 t^{-\frac1{(m+1)}} f\left(k t^{-\frac1{(m+1)}}\right) dk
=
t^{\frac2{(m+1)}}
 \int \xi^2 f(\xi) d \xi,
\end{equation}
so
$\sigma(t) \sim t^{\frac2{(m+1)}}$ which is usually refered to as a sub-diffusive spreading.
 In particular, for the four-wave systems
we have $\sigma(t) \sim t^{1/2}$
and for the six-wave systems, respectively, $\sigma(t) \sim t^{1/3}$.

Plugging \eqref{SelfSimilar} into \eqref{PorousMedium2} yields the following equation for $f$,
\begin{equation}
\label{Fequation}
(m+1) (f^m)'' \ + \  \xi f' \ + f =0.
\end{equation}
Integrating this equation once, we get
\begin{equation}
\label{Fequation1}
(m+1) (f^m)' \ + \  \xi  f =C,
\end{equation}
where $C=$const.
Let us consider first the case with $C\ne0$: according to the above
equation $(f^m)' < C/(m+1)$, so the solution has a sharp front at $\xi=
\xi^*$ such that $f(\xi^*) =0$. (This front will be on the right boundary of the solution
for $C>0$ and on the left boundary for $C<0$.)
But the current $J = - \partial_k N^m =  t^{-1} (f^m)' $ will remain
finite at  $\xi=
\xi^*$, which means that there is a moving sink of particles at $k=\xi^* t^{-a}$.
Thus, the solutions with $C\ne0$ are unphysical.

Thus, we put $C=0$ and solve equation \eqref{Fequation1} directly, which gives \cite{tuck76}
\begin{equation}
\label{SpecialSolution}
f(\xi) \ = \ \left[\frac{(m-1)}{2m(m+1)}(\xi^{*2}-\xi^2 )
\right]^\frac{1}{m-1}
\end{equation}
with some constant $\xi^*$ which, again, corresponds to a sharp moving boundary of the solution.
Considering negative $k$ one can see that the solution remains in the same
form \eqref{SpecialSolution}. Thus, including both negative and positive $k$, we have a
solution with the shape of a droplet with sharp boundaries expanding infinitely on a flat surface.

This shows that one can construct a solution to the porous medium equation, that spreads infinitely for large times.
Moreover, the theory of porous medium equations says that such a self-similar solution  is stable,
and that it is an attractor for all solutions with arbitrary localised initial conditions.
This result apply to any $m>1$, including
$m=3$ (four-wave regime considered in this
paper) and $m=5$ (six-wave regime outlined in the next section).
 Therefore this answers the ``weak turbulence'' question \eqref{Question} positively.

%
%
%

\section{Six-wave regime}

Here, we will present a speculative discussion of the cases when the four-wave interaction
considered in this paper may become ineffective and subdominant to a higher-order process -- the six-wave
regime. Let $\omega$ (the maximal strength of the  potentials)
be small compered to $k^2$ ($k$ being the typical wave momentum), and $k \ll 1$. In  the limit we get the  continuous NLS without potentials, which is integrable and, therefore, the resonant interactions of every order are null. Small deviation from the limit of zero potentials and continuous space result in small nonintegrability and, therefore, in activation of wave resonances. However, the four-wave process is still null, because  the four-wave frequency and momentum resonant conditions cannot be satisfied in 1D for the dispersion relations $E=k^2$, and this property cannot be removed by small perturbations.   Due to the $U(1)$ symmetry of the problem, the odd-order resonant processes are absent, and the leading order process is expected to be six-wave.

Based on analogy with equation \eqref{Kinetic1},
we can conjecture that the six-wave kinetic equation in this case will take form
\begin{equation}
\label{Kinetic6}\begin{aligned}
\dot{N}_j \ = \iiint_{\mathbb{R}^3}L(l-j,m-j,n-j, p-j, q-j)
N_jN_lN_mN_n N_p N_q \\
(N_j^{-1}+N_l^{-1} +N_m^{-1} -N_n^{-1}- N_p^{-1} -N_q^{-1})dmdndldpdq,\end{aligned}
\end{equation}
where
\begin{equation}
\label{Lernel}
L(l-j,m-j,n-j, p-j,q-j) \ := \  \epsilon^4  \langle |W_{jlm}^{npq}|^2  \delta (E_{jlm}^{npq}) \rangle
\end{equation}
with $W_{jlm}^{npq}$ being an interaction coefficient which is probably quite complicated, since
obtaining it should involve a canonical transformation removing the cubic nonlinearity from the
dynamical equation (as it is usual when the four-wave process is absent).
It is natural to expect that in this case the energy and momentum will be
approximately conserved and the kernel $L$ will be weakly localised near values
$j\sim l \sim m \sim n \sim p \sim q$. (The closer we are to the limit of the zero
potentials and continuous medium, the better the energy/momentum conservation,
and the weaker the localisation in the physical space.)

\section{Summary and discussion}

In the present paper, we developed a wave turbulence theory for description
of weak excitations in the model described by the discrete one dimensional NLS equation
with random potentials \eqref{NLSRandom1}.
We systematically derived a four-wave kinetic equation
 \eqref{Kinetic0} and its continuous version  \eqref{Kinetic1}.
 From the latter, we derived a porous medium equation   \eqref{PorousMedium}
 for the cases when the linear mode
 localisation length is less than the characteristic length of the wave spectrum variation.
 Such porous medium equation was previously suggested for the discrete NLS model
 in \cite{Kolovsky10,Flach2015}, and in the present paper we
 elevate the status of this equation as fully justified via a systematic wave turbulence derivation.
 Further, we presented a speculative argument about the conditions when the four-wave regime
 is replaced by a six-wave process described by the kinetic equation  \eqref{Kinetic6}
 and (in the case of slow spatial variations of the wave spectra) by the $m=5$ version
 of the porous medium equation  \eqref{PorousMedium2}.

 Analysing stationary solutions of the porous medium equation, we have obtained an effective
 Ohm's law -- a nonlinear current-voltage relation indicating that weak nonlinearity makes
 the lattice a conductor. This is one of the ways to the localisation is broken.

 Another, more traditional way to characterise de-localisation by nonlinearity is
 to study the self-similar solution of the porous medium equation  \eqref{PorousMedium2}.
 For any $m>1$, the self-similar spreading appears to be sub-diffusive:
 regime $\sigma(t) \sim t^{1/2} $ is realised for the four-wave case ($m=2$)
 whereas the six-wave regime ($m=3$) leads to $\sigma(t) \sim t^{1/3} $.
 Numerical experiments  of \cite{Kolovsky10}
 reported observation of the $\sigma(t) \sim t^{1/2} $ regime, whereas
 other numerical experiments
  \cite{pikovsky2008destruction,flach2009universal,skokos2009delocalization,Flach2015}
  reported $\sigma(t) \sim t^{1/2} $. Interestingly, a $\sigma(t) \sim t^{1/3} $ to $\sigma(t) \sim t^{1/2} $
  was observed in
    \cite{flach2009universal,skokos2009delocalization,Flach2015}
    when the wave phases where artificially scrambled.
    A  connection between the degree of phase randomness and
    realisability of either four-wave or six-wave dynamics remains to be understood.
    It is quite possible that the  phases become random only after passing
    to variables obtained via the canonical transformation (required for
    deriving the six-wave kinetic equation) and that the phases of the
    original variables are correlated.

  ~~ \\{\bf Acknowledgements.}
  S. Nazarenko is supported by the following grants: "Photons for Quantum Simulation" (EC H2020-FETFLAG-2018-03, project 820392), "Hydrodynamic Approach to Light Turbulence" (EC H2020-MSCA-RISE-2018, project 823937), and Chaire D’Excellence IDEX, Universit\'e de la C\^ote d'Azur, France.
A. Soffer is partially supported by
NSF grant DMS 1600749  and NSFC 11671163.
M.-B Tran is partially supported by NSF Grants DMS-1814149 and DMS-1854453, and a Sam Taylor Fellowship.
\bibliographystyle{plain}

\bibliography{QuantumBoltzmann}

\def\cprime{$'$}
\begin{thebibliography}{10}

\bibitem{balk1990physical}
A.~M. Balk and S.~V. Nazarenko.
\newblock Physical realizability of anisotropic weak-turbulence kolmogorov
  spectra.
\newblock {\em Sov. Phys. JETP}, 70:1031--1041, 1990.

\bibitem{berman2005fermi}
G.~P. Berman and F.~M. Izrailev.
\newblock The fermi--pasta--ulam problem: fifty years of progress.
\newblock {\em Chaos: An Interdisciplinary Journal of Nonlinear Science},
  15(1):015104, 2005.

\bibitem{bourgain2009mathematical}
J.~Bourgain, C.~E. Kenig, and S.~Klainerman.
\newblock {\em Mathematical Aspects of Nonlinear Dispersive Equations
  (AM-163)}.
\newblock Princeton University Press, 2009.

\bibitem{campbell2005introduction}
D.~K Campbell, P.~Rosenau, and G.~M. Zaslavsky.
\newblock Introduction: The fermi--pasta--ulam problem: the first fifty years.
\newblock {\em Chaos: An Interdisciplinary Journal of Nonlinear Science},
  15(1):015101, 2005.

\bibitem{CHOI2004}
Yeontaek Choi, Yuri~V. Lvov, and Sergey Nazarenko.
\newblock Probability densities and preservation of randomness in wave
  turbulence.
\newblock {\em Physics Letters A}, 332(3):230 -- 238, 2004.

\bibitem{CHOI2005a}
Yeontaek Choi, Yuri~V. Lvov, and Sergey Nazarenko.
\newblock Joint statistics of amplitudes and phases in wave turbulence.
\newblock {\em Physica D: Nonlinear Phenomena}, 201(1):121 -- 149, 2005.

\bibitem{CHOI2005}
Yeontaek Choi, Yuri~V. Lvov, Sergey Nazarenko, and Boris Pokorni.
\newblock Anomalous probability of large amplitudes in wave turbulence.
\newblock {\em Physics Letters A}, 339(3):361 -- 369, 2005.

\bibitem{CraciunBinh}
G.~Craciun and M.-B. Tran.
\newblock A reaction network approach to the convergence to equilibrium of
  quantum boltzmann equations for bose gases.
\newblock {\em arXiv preprint arXiv:1608.05438}, 2016.

\bibitem{FaouGermainHani:TWN:2016}
E.~Faou, P.~Germain, and Z.~Hani.
\newblock The weakly nonlinear large-box limit of the 2{D} cubic nonlinear
  {S}chr\"odinger equation.
\newblock {\em J. Amer. Math. Soc.}, 29(4):915--982, 2016.

\bibitem{fishman2009perturbation}
S.~Fishman, Y.~Krivolapov, and A.~Soffer.
\newblock Perturbation theory for the nonlinear schr{\"o}dinger equation with a
  random potential.
\newblock {\em Nonlinearity}, 22(12):2861, 2009.

\bibitem{fishman2012nonlinear}
S.~Fishman, Y.~Krivolapov, and A.~Soffer.
\newblock The nonlinear schr{\"o}dinger equation with a random potential:
  results and puzzles.
\newblock {\em Nonlinearity}, 25(4):R53, 2012.

\bibitem{Flach2015}
S.~Flach.
\newblock {\em Nonlinear Lattice Waves in Random Potentials}, pages 1--48.
\newblock Springer International Publishing, 2015.

\bibitem{Flach:SNS:2016}
S~Flach.
\newblock Spreading, nonergodicity, and selftrapping: a puzzle of interacting
  disordered lattice waves.
\newblock In {\em Nonlinear dynamics: materials, theory and experiments},
  volume 173 of {\em Springer Proc. Phys.}, pages 45--57. Springer, Cham, 2016.

\bibitem{flach2009universal}
S.~Flach, D.~Krimer, and Ch. Skokos.
\newblock Universal spreading of wave packets in disordered nonlinear systems.
\newblock {\em Physical Review Letters}, 102(2):024101, 2009.

\bibitem{galtier2000weak}
S.~Galtier, S.~Nazarenko, A.~C. Newell, and A.~Pouquet.
\newblock A weak turbulence theory for incompressible magnetohydrodynamics.
\newblock {\em Journal of Plasma Physics}, 63(5):447--488, 2000.

\bibitem{germain2017optimal}
P.~Germain, A.~D. Ionescu, and M.-B. Tran.
\newblock Optimal local well-posedness theory for the kinetic wave equation.
\newblock {\em arXiv preprint arXiv:1711.05587}, 2017.

\bibitem{hani2014long}
Z.~Hani.
\newblock Long-time instability and unbounded sobolev orbits for some periodic
  nonlinear schr{\"o}dinger equations.
\newblock {\em Archive for Rational Mechanics and Analysis}, 211(3):929--964,
  2014.

\bibitem{JinBinh}
S.~Jin and M.-B. Tran.
\newblock Quantum hydrodynamic approximations to the finite temperature trapped
  bose gases.
\newblock {\em Physica D: Nonlinear Phenomena}, 380-381:45--57, 1 October 2018.

\bibitem{Kolovsky10}
A.R. Kolovsky, E.A. Gomez, and H.J. Korsch.
\newblock Bose-einstein condensates on tilted lattices: Coherent, chaotic, and
  subdiffusive dynamics.
\newblock {\em Phys. Rev. A}, 81:025603, 2010.

\bibitem{krivolapov2010numerical}
Y.~Krivolapov, S.~Fishman, and A.~Soffer.
\newblock A numerical and symbolical approximation of the nonlinear anderson
  model.
\newblock {\em New Journal of Physics}, 12(6):063035, 2010.

\bibitem{lvov2004noisy}
Y.~V. Lvov and S.~Nazarenko.
\newblock Noisy spectra, long correlations, and intermittency in wave
  turbulence.
\newblock {\em Physical Review E}, 69(6):066608, 2004.

\bibitem{mithun2018weakly}
T.~Mithun, Y.~Kati, C.~Danieli, and S.~Flach.
\newblock Weakly nonergodic dynamics in the gross-pitaevskii lattice.
\newblock {\em Physical review letters}, 120(18):184101, 2018.

\bibitem{Nazarenko:2011:WT}
S.~Nazarenko.
\newblock {\em Wave turbulence}, volume 825 of {\em Lecture Notes in Physics}.
\newblock Springer, Heidelberg, 2011.

\bibitem{NewellNazarenko:2001:WTA}
A.~C. Newell, S.~Nazarenko, and L.~Biven.
\newblock Wave turbulence and intermittency.
\newblock {\em Phys. D}, 152/153:520--550, 2001.
\newblock Advances in nonlinear mathematics and science.

\bibitem{pikovsky2008destruction}
A.~S. Pikovsky and D.~L. Shepelyansky.
\newblock Destruction of anderson localization by a weak nonlinearity.
\newblock {\em Physical review letters}, 100(9):094101, 2008.

\bibitem{PomeauBinh}
Y.~Pomeau and M.-B. Tran.
\newblock Statistical physics of non equilibrium quantum phenomena.
\newblock {\em Book}, 2018.

\bibitem{ReichlTran}
L.~E. Reichl and M.-B. Tran.
\newblock A kinetic equation for ultra-low temperature bose--einstein
  condensates.
\newblock {\em Journal of Physics A: Mathematical and Theoretical},
  52(6):063001, 2019.

\bibitem{rivkind2011eigenvalue}
A.~Rivkind, Y.~Krivolapov, S.~Fishman, and A.~Soffer.
\newblock Eigenvalue repulsion estimates and some applications for the
  one-dimensional anderson model.
\newblock {\em Journal of Physics A: Mathematical and Theoretical},
  44(30):305206, 2011.

\bibitem{skokos2009delocalization}
Ch. Skokos, D.~Krimer, S.~Komineas, and S.~Flach.
\newblock Delocalization of wave packets in disordered nonlinear chains.
\newblock {\em Physical Review E}, 79(5):056211, 2009.

\bibitem{SofferBinh2}
A.~Soffer and M.-B. Tran.
\newblock On coupling kinetic and schrodinger equations.
\newblock {\em Journal of Differential Equations}, 265 (5):2243--2279, 2018.

\bibitem{SofferBinh1}
A.~Soffer and M.-B. Tran.
\newblock On the dynamics of finite temperature trapped bose gases.
\newblock {\em Advances in Mathematics}, 325:533--607, 2018.

\bibitem{tuck76}
B.~Tuck.
\newblock Some explicit solutions to the non-linear diffusion equation.
\newblock {\em J. Phys. D}, 9:1559, 1976.

\bibitem{vakulchyk2019wave}
I.~Vakulchyk, M.~V. Fistul, and S.~Flach.
\newblock Wave packet spreading with disordered nonlinear discrete-time quantum
  walks.
\newblock {\em Physical review letters}, 122(4):040501, 2019.

\bibitem{Vazquez:TPM:2007}
J.~L. V\'azquez.
\newblock {\em The porous medium equation}.
\newblock Oxford Mathematical Monographs. The Clarendon Press, Oxford
  University Press, Oxford, 2007.
\newblock Mathematical theory.

\bibitem{wang2009long}
W.~M. Wang and Z.~Zhang.
\newblock Long time anderson localization for the nonlinear random
  schr{\"o}dinger equation.
\newblock {\em Journal of Statistical Physics}, 134(5-6):953--968, 2009.

\bibitem{zakharov2012kolmogorov}
V.~E. Zakharov, V.~S. Lvov, and G.~Falkovich.
\newblock {\em Kolmogorov spectra of turbulence I: Wave turbulence}.
\newblock Springer Science \& Business Media, 2012.

\end{thebibliography}

\end{document}